# Evidence for single variant in altermagnetic RuO$_2$(101) thin films

Cong He,[1,§,#] Zhenchao Wen,[1,§,]* Jun Okabayashi,[2,]* Yoshio Miura,[1,3]* Tianyi Ma,[1] Tadakatsu Ohkubo,[1] Takeshi Seki,[4,5] Hiroaki Sukegawa,[1] and Seiji Mitani[1]

[1]National Institute for Materials Science (NIMS), Tsukuba 305-0047, Japan

[2]Research Center for Spectrochemistry, The University of Tokyo, Bunkyo, Tokyo 113-0033, Japan

[3]Faculty of Electrical Engineering and Electronics, Kyoto Institute of Technology, Kyoto 606-8585, Japan

[4]Institute for Materials Research, Tohoku University, Sendai 980-8577, Japan

[5]Center for Science and Innovation in Spintronics, Tohoku University, Sendai 980-8577, Japan

[§]These authors contributed equally to this work. [#]Present address: Hunan University, Changsha 410082, China. * Wen.Zhenchao@nims.go.jp, jun@chem.s.u-tokyo.ac.jp, miura@kit.ac.jp

**Abstract**

Altermagnetism presents intriguing possibilities for spintronic devices due to its unique combination of strong spin-splitting and zero net magnetization. However, realizing its full potential hinges on fabricating single-variant altermagnetic thin films. In this work, we present definitive evidence for formation of single-variant altermagnetic RuO$_2$(101) thin films with fully epitaxial growth on Al$_2$O$_3$(1$\bar{1}$02) $r$-plane substrates, confirmed through rigorous structural analyses using X-ray diffraction, atomic-resolution transmission electron microscopy and X-ray magnetic linear dichroism. The mutual correspondence of the occupancy of oxygen atoms on the surfaces of RuO$_2$(101)[010] and Al$_2$O$_3$(1$\bar{1}$02)[11$\bar{2}$0] plays a decisive role in the formation of the single-variant RuO$_2$, which is also supported by our first-principles density functional theory calculations. We further observed spin-splitting magnetoresistance in the single-variant RuO$_2$(101)/CoFeB bilayers, highlighting the characteristic effect of single variant on spin transport. The demonstration of single-variant RuO$_2$(101) films marks a significant advancement in the field of altermagnetism and paves the way for exploring their potential applications.

**INTRODUCTION**

Altermagnetism represents a unique phase of magnetism, distinct from the conventional characteristics of ferromagnetism and antiferromagnetism.[1–4] Unlike traditional magnetic materials, altermagnets exhibit large spin splitting but maintain zero net magnetization due to the combined protection of spin and real-space group symmetry.[5–12] In essence, they exhibit the strong spin polarization typical of ferromagnets, while retaining the antiparallel magnetic ordering characteristic of antiferromagnets. This intriguing combination of ferromagnetic-like spin polarization and antiferromagnetic order holds immense potential for spintronic applications, such as spin-orbit torque (SOT) and magnetoresistive devices.[13–19] The number of potential altermagnetic materials is steadily growing, with candidates ranging from insulators to metals.[1,9,12,20–23] Among them, tetragonal RuO$_2$ has emerged as a particularly exciting altermagnet. It is a conductive rutile oxide and was considered as paramagnet, but later proved to be itinerant antiferromagnet,[24,25] and more recently demonstrated to have strong time-reversal symmetry breaking in the band structure of altermagnetic RuO$_2$.[12] A prominent anomalous Hall effect was reported in RuO$_2$ thin films due to the altermagnetic phase.[26,27] Néel spin currents and tunneling magnetoresistance effect have also been proposed in RuO$_2$/TiO$_2$/RuO$_2$(001)[17,19] and RuO$_2$/TiO$_2$/CrO$_2$(110)[28] heterostructures. Further, it was demonstrated that RuO$_2$ enabled efficient generation of spin currents due to its spin splitting effect.[13–16] The efficiencies of the spin current generation and the spin direction of the resulting spin currents are strongly dependent on the crystallographic orientation of the RuO$_2$ tetragonal lattice as well as the Néel vector orientation.

Among the various crystal facets of RuO$_2$, the (101)-oriented RuO$_2$ film stands out as a promising platform for spintronics applications, as it has recently been used to achieve SOT-induced magnetization switching without applying any external magnetic field.[15] In RuO$_2$(101) films, thanks to the tilting of the Néel vector with respect to the film surface, tilted spin currents are generated when the charge current is applied along the RuO$_2$[010] direction,[13,15,16] which is crucial to achieve magnetic-field-free magnetization switching. Figure 1a shows the illustration of crystal structure of the altermagnetic RuO$_2$ where the Néel vector is parallel to the [001] direction. The (101) atomic plane is indicated by a grey color and it has an angle of ~55° with respect to the Néel vector. We name it variant A. In crystallography, RuO$_2$(101) has an equivalent variant B, i.e., RuO$_2$($\bar{1}$01), as shown in Fig. 1b. However, the orientations of the Néel vectors of the two variants are not at all the same, but rather in mirror symmetry. Because the (101) and ($\bar{1}$01) lattice planes are equivalent and have the same lattice constants, two variants with different oriented Néel vectors would exist in the film, which will have important implications for spin transport. The schematic of the prototype band structure of RuO$_2$ is illustrated in Fig. 1c in which the spin-splitting Fermi surface is clearly shown for up-spin and down-spin. Once the presence of two variants in a thin film, it complicates materials design and hinders precise control over spin transport properties. Thus, a single variant RuO$_2$ is essential to unlock the full potential of spin splitting nature to revolutionize spintronics and catalyze the design of novel devices. To date, there has been no direct evidence confirming the existence of a single variant of RuO$_2$(101).

In this work, we fabricated RuO$_2$(101) epitaxial thin films on Al$_2$O$_3$(1$\bar{1}$02) *r*-plane substrates and utilized exhaustive structural analyses, including X-ray diffraction (XRD), atomic-resolution scanning transmission electron microscopy (STEM) and X-ray magnetic linear dichroism (XMLD) techniques, to provide direct evidence for the existence of a single variant in the RuO$_2$(101) films.

We discovered that the precise matching of oxygen atom arrangements on the surfaces of RuO$_2$(101) and Al$_2$O$_3$(1$\bar{1}$02) is crucial in determining the formation of the single-variant RuO$_2$(101). First-principles density functional theory (DFT) calculations also proved its stability. XMLD can distinguish compensated magnetic structures and the formation of charge quadrupoles, undetectable by X-ray magnetic circular dichroism (XMCD). We performed XMLD with angular dependence to detect the Néel vector direction by probing the charge distributions coupled with linearly polarized synchrotron beams, which revealed a single-variant altermagnetic RuO$_2$(101) film with finite charge quadrupole and spin moment, in agreement with our theoretical calculations. We further demonstrated the spin-splitting magnetoresistance (SSMR) in the single-variant altermagnetic RuO$_2$(101)/CoFeB heterostructures associated with the distinctive single variant and titled spin current.

**RESULTS AND DISCUSSION**

**A. Structural and electronic transport characterizations**

Figure 1d shows the out-of-plane XRD scan results for the films deposited on the *r*-plane sapphire substrate at 300 °C and subsequently annealed at $T_a$ = 300, 500, 600, and 700 °C. Apart from the substrate peaks, distinct diffraction peaks along the [101] orientation were observed, indicating favorable growth in this orientation. Note that here, the [101] can also be equivalent to the [$\bar{1}$01] orientation considering the crystallographic symmetry. Figure 1e indicates the variation in full width at half maximum (FWHM) and diffraction intensity of 101 and 202 peaks to assess film crystallinity with annealing temperature. It is observed that higher annealing temperatures enhance crystallinity, evidenced by reduced FWHM and increased intensity. The interplanar lattice spacing of RuO$_2$(101) in the films is measured to be ~0.2538 nm, slightly smaller than its bulk counterpart (0.2555 nm). In order to distinguish whether the film contains variant A or B or both, we further performed in-plane XRD measurements. The detailed description of the measurement setup and method are shown in Supplementary Materials. We firstly set up the configuration of X-ray incidence and reflection for RuO$_2$(002) plane by rotating the sample to $\chi= 34.08º$, $2\theta = 59.62º$, then carried out the in-plane $\varphi$ scan. The X-ray incident slit used in the measurement is 10 mm in size, which is large enough to obtain information about the crystal structure of the entire thin film. The result of in-plane XRD scan for RuO$_2$(002) lattice plane of the RuO$_2$(101) thin film is shown in Fig. 1f. A single peak was clearly observed, which indicates a single-variant film. Note that we would see two peaks for the case of two variants by rotating the sample one full 360º circle in the plane of the sample surface because of the reverse variant RuO$_2$($\bar{1}$01). In order to further confirm this single variant feature, we also performed the in-plane XRD scan measurements for RuO$_2$(10$\bar{1}$) and (200) lattice planes at $\chi = 68.85º$, $2\theta = 35.29º$, and $\chi = 55.28º$, $2\theta = 40.47º$, respectively. The XRD patterns are shown in Figs. 1g and 1h. Both cases show a characteristic single peak. The above results indicate that only variant A exists in the RuO$_2$(101) thin film deposited on an Al$_2$O$_3$(1$\bar{1}$02) substrate. Figure 1i displays the RHEED pattern along the RuO$_2$[10$\bar{1}$] direction, showing prominent streaks indicative of high-quality film preparation with a very flat surface morphology. We also investigated RuO$_2$ thin films grown on various substrates under different conditions and found that films on Al$_2$O$_3$(1$\bar{1}$02) exhibited the highest conductivity of $2.1 \times 10^6$ $\Omega^{-1}$m$^{-1}$ achieved at $T_a = 600$ °C. Additionally, Fig. 1j shows the analysis of temperature-dependent resistivity which reveals a Néel temperature around 390 K for films, indicating the transition from antiferromagnetic to paramagnetic states. More structure and electronic transport characterizations of the samples are in Supplementary Materials.

We further employed high angle annular dark-field STEM (HAADF-STEM) for atomic-resolution structural analysis for the single variant RuO$_2$(101) film. The geometric relationship among the Néel vectors, $RuO_2(101)$ and $(\bar{1}01)$ planes are shown in Fig. 2a where the angle between the Néel vector and (101) or $(\bar{1}01)$ plane is ~55°. If two types of variants were both formed in the $RuO_2$ film, two sets of Néel vectors would be obtained among which one Néel vector deviates ~70° from the other. The HAADF-STEM image in Fig. 2b shows the atomic-resolution cross-sectional microstructure of the altermagnetic $RuO_2$ film grown on the *r*-plane $Al_2O_3$ substrate, with the viewing direction being parallel to the $RuO_2$[010] direction. A much larger scale of HAADF-STEM image is shown in Supplementary Materials. The corresponding nanobeam electron diffraction (NBED) patterns collected from the film and substrate are shown in the upper-right and lower-right corners, respectively, which is also consistent with the epitaxial relationship by the XRD results. Figure 2c displays the enlarged HAADF-STEM image at the interface between the $Al_2O_3$ substrate and the $RuO_2$ film, where a good lattice matching is readily detected. The orange line is the trace direction of its Néel vector, having an angle of ~55° relative to the film plane. The specific atomic-matching scenario at the interface of $Al_2O_3(1\bar{1}02)/RuO_2(101)$ is shown in the schematic diagram in Fig. 2d, which is consistent with the experimental result in Fig. 2c. This can be attributed to the fact that the $RuO_2(101)$ plane, rather than the $RuO_2(\bar{1}01)$ (see the schematic of the cross-sectional lattice matching between $RuO_2(\bar{1}01)$ and $Al_2O_3(1\bar{1}02)$ in Supplementary Materials), was formed on the surface of $Al_2O_3(1\bar{1}02)$ substrate. The in-plane atomic lattice matching among $Al_2O_3(1\bar{1}02)$, $RuO_2(101)$ and $RuO_2(\bar{1}01)$ is shown in Figs. 2e-2g, where only a single atomic plane of Ru atoms, and two layers of O atoms close to the Ru atomic layer are shown. Our crystallographic analysis clearly reveals that the in-plane atomic arrangement of $RuO_2(101)$ is nearly same as that of $Al_2O_3(1\bar{1}02)$. The O atomic position in the $RuO_2(\bar{1}01)$ in-plane atomic arrangement, however, is different from that in the $Al_2O_3(1\bar{1}02)$ case. To be specific, the two layers of O atoms have reversed their positions, i.e., the previous O layer above the Ru atoms now becomes the O layer below Ru, as highlighted by the dashed blue circles in Figs. 2e-2g. Therefore, on the surface of $Al_2O_3(1\bar{1}02)$ substrate, the growth of $RuO_2(101)$ plane is preferable. We also conducted first-principles DFT calculations to clarify the stability of the $Al_2O_3(1\bar{1}02)$ surface (See Supplemental Materials). The calculations reveal that the substrate surface terminated with Al and O ($z = +0.5$) atoms (Fig. 2e) is the most stable. We further calculated the stability of the $RuO_2(101)$ and $RuO_2(\bar{1}01)$ variants by stacking $RuO_2(101)$ and $RuO_2(\bar{1}01)$ cells on the $Al_2O_3(1\bar{1}02)$ surface, corresponding to Fig. 2f [$Al_2O_3(1\bar{1}02)/RuO_2(101)$] and Fig. 2g [$Al_2O_3(1\bar{1}02)/RuO_2(\bar{1}01)$]. It is found that $Al_2O_3(1\bar{1}02)/RuO_2(101)$ has a lower total formation energy density by 1.2 $J/m^2$ compared to $Al_2O_3(1\bar{1}02)/RuO_2(\bar{1}01)$. This result confirms that the $Al_2O_3(1\bar{1}02)/RuO_2(101)$ is the more favorable configuration, aligning with experimental observations. We note that the possibility of mixed surface terminations, due to thermodynamic competition and kinetic factors, may affect the stabilization of $RuO_2(101)$ versus $RuO_2(\bar{1}01)$ variants. Once two variants exist, one would detect two peaks in the in-plane XRD $\varphi$ scan (Figs. 1f-1h). Therefore, surface treatment of the substrate and precise control of growth conditions are crucial for achieving single-variant $RuO_2(101)$ thin films.

**B. Detecting magnetic states of altermagnetic $RuO_2(101)$ thin films by angular-dependent XMLD**

Figure 3 shows the X-ray absorption spectroscopy (XAS), XMCD, and XMLD at Ru $M$-edges. In the XAS, both 3$p$ to 4$d$ and 3$p$ to 5$s$ absorption peaks with spin-orbit splitting of $3p_{3/2}$ ($M_3$) and $3p_{1/2}$ ($M_2$) levels are observed. Although the cross-sections in $M$-edge absorption are smaller than those in $L$-edges, finite signals of XAS and differences between linear polarized beams along horizontal and vertical directions can be detected overlapping with linear background components of negative slope as displayed in Fig. 3a. The satellite peaks at 475 and 500 eV correspond to the 3p to 5s absorption. In the case of circularly polarized beams, no XMCD signals are detected as shown in Fig. 3b because of completely compensated antiferromagnetic order, which is different from the previous reports of Ru M-edge XMCD in Ru-based ferromagnetic compounds.[29–31] Next, the XMLD signals detected by linearly polarized beams are displayed with sample angular dependence. Since the electric-field components of linearly polarized beams couple with those of magnetization, it depends on the rotation angles along in-plane between the beam and the antiferromagnetic Néel vectors. Figures 3c-h display the angular dependence of XMLD, which clearly shows the difference between positive and negative angles. Here, the angle between the sample surface normal and beam direction is defined as $\theta$, and the angle that the sample is rotated in the plane of film surface is $\varphi$. The geometry of $\theta = 0°$ corresponds to the normal incidence case in Figs. 3d and 3g. Clear differential XMLD line shapes are also observed. The XMLD signal intensities are most enhanced in the case of $\theta = -55°$ tilt and gradually suppressed at 0° and +35°, which can determine the Néel vector orientation along the [001] direction and the $3z^2 - r^2$ orbital states are coupled mainly with the linearly polarized beam. To confirm these facts, the sample is rotated at $\varphi = 90°$, the spins are orthogonal, and canted components can couple with the linearly polarized beams as shown in the insets of Figs. 3f-h. With the sample rotation, the XMLD intensities are almost identical because of canting along the orthogonal direction. Since the beam spot size is on the order of micrometers, systematic changes in XMLD result from the formation of single variants as probed by XRD. Therefore, these results indicate that (i) the linear dichroism components arising from lattice symmetry are negligible, (ii) the antiferromagnetic ordering in the Ru sites with finite magnetic moments is endured, and (iii) the Néel vector orientation of $RuO_2$ is aligned along the [001] direction. Note that since the XMLD signals may include the structural information as X-ray linear dichroism (XLD), we performed the temperature-dependent XMLD measurements up to 400 K, higher than the Néel temperature of $RuO_2$, to eliminate the XLD contribution, as shown in Supplementary Materials. The results show a monotonic decrease in XMLD signal amplitude with increasing temperature and a significant suppression at 400 K. The weak spectrum at 400 K above the Néel temperature can reflect the structural contribution. To our knowledge, the structural component of XMLD (i.e., XLD) is generally temperature-independent unless a structural phase transition occurs. Since our temperature-dependent resistivity measurements indicate no such structural transitions, the significant temperature-dependent XMLD signal supports the magnetic origin and provides the evidence of antiferromagnetic ordering in our $RuO_2$(101) samples. Although the magnetism in the $RuO_2$ is still controversial in recent experimental reports,[32,33] the discrepancies may be due to sample-to-sample variations, especially differences in growth conditions, Ru or O vacancies, and structural uniformity.

The XMLD spectral line shape analyses provide the quadrupole components ($Q_{zz}$) of the Ru sites using an XMLD magneto-optical sum rule.[34–36] Since the $Q_{zz}$ represents the degree of magnetically polarized charge anisotropy at the atomic sites, the electronic structure of the distorted Ru sites can be probed. The integrals of the XMLD spectra, excluding the 3$p$ to 5$s$ absorption contributions, can monitor the values of $Q_{zz}$, where asymmetric $L_3$ and $L_2$ line shapes result in the finite converged values of $Q_{zz}$ (See Supplemental Materials). In the XMLD shown in Fig. 3c, positive converged values are estimated, suggesting that the quadrupoles elongating along the axial directions as a function of $3z^2 - r^2$ are predominant. The Ru units are surrounded by octahedral oxygen sites with distortion along the axial oxygen atoms and are alternately aligned with the rotation of 90°. Since the XMLD (XLD) probes the sums of two kinds of Ru units, the charge distortions of up and down spin sites are almost cancelled as suggested by the theoretical calculations shown in Fig. 4, and XMLD in $RuO_2$ probes only magnetically polarized information.

Therefore, the systematic changes of angle-dependent XMLD suggest the single variant feature in the altermagnetic RuO$_2$(101) thin film.

## C. Magnetic anisotropy calculations for the altermagnetic RuO$_2$

The magneto-crystalline anisotropy energy (MAE) between the [001] and [100] magnetization directions in altermagnetic RuO$_2$ was investigated by the first-principles DFT calculations. We found that RuO$_2$ has the [001] easy axis with respect to the [100] magnetization direction, which corresponds to the MAE energy of 1.44×10$^7$ J/m$^3$. Here, the spin magnetic moments are estimated to be 1.12 $\mu_B$ for each Ru site and antiferromagnetically aligned to compensate for the total net spins. The MAE value is comparable to that of $L1_0$-FePt based on the DFT calculations, implying that altermagnetic RuO$_2$ has a strong MAE along the [001] direction. To clarify the origin of the MAE of RuO$_2$, a second-order perturbation analysis in terms of the spin-orbit interaction was performed. The MAE of each Ru atomic site $I$ (Ru$_1$, Ru$_2$) can be formulated as

$$E_{MAE}^I = \pm \frac{\xi_I}{4}(\langle L_z \rangle - \langle L_x \rangle) \pm \frac{21}{8}\frac{\xi_I^2}{\Delta_{ex}^I}(-\langle Q_{zz}S_z \rangle), \qquad (1)$$

where the first term $\langle L_z \rangle - \langle L_x \rangle$ and the second term $-\langle Q_{zz}S_z \rangle$ represent the expected values of orbital moment anisotropy and spin quadrupole moment, respectively. $\xi_I$ and $\Delta_{ex}^I$ are the spin-orbit constant and exchange splitting of each atomic site, respectively.[43] Note that the plus (+) and minus (−) signs depend on positive and negative values of the spin moment $\langle S_z \rangle$, since the direction of the spin quantum axis at each atomic site depends on the direction of the local spin moment. Figures 4a and 4b illustrate the Ru and O sites in the units of RuO$_2$ used in this calculations with the quadrupole and spin moments. The contribution to the MAE of the first term (spin-conserving) and the second term (spin-flipping) within the Ru sites are shown in Fig. 4c. The positive and negative values indicate the [001] and [100] easy axes of the magnetization, respectively. It is found that the Ru atoms mainly contribute to the MAE along the [001] direction, while the contribution of oxygen to the MAE is negligible. In particular, the spin-flip term contributes more than twice as much to the MAE as the spin-conserving term. This suggests that the spin quadrupole of the Ru sites has a cigar-type distribution along the [001] direction as shown in Fig. 4a, which mainly contributes to the MAE of $RuO_2$. Note that the spin-conserving term is also positive. The fact that both spin-flip and spin-conserving terms contribute to the MAE is unique and identical for altermagnetism-based antiferromagnets. Figure 4d shows the orbital moment anisotropy (difference between the [001] and [100] directions) and the spin quadrupole moment ($-\langle Q_{zz}S_z \rangle$) for the Ru atom. It is clear that the quadrupole term is larger than the orbital moment anisotropy, which is consistent with the dominant contribution of the spin-flip term in Fig. 4c. The band dispersions and MAE contributions along the high symmetric lines of rutile-type $RuO_2$ ($P4_2/mnm$) are shown in Figs. 4e and 4f. Looking at the spin quadrupole moments from the $z$-axis ([001] direction), one would expect anisotropy in the [110] and [$\bar{1}$10] directions due to the anisotropic oxygen bonding in the rutile structure. However, the in-plane anisotropies are counteracted by the antiferromagnetic magnetic structure of the $Ru_1$ and $Ru_2$ sites. Thus, there remains the cigar-type spin quadrupole moments in the $z$-axis caused by the distorted tetragonal structure of $RuO_2$. This cigar-type spin quadrupole moment of Ru has been observed experimentally and is considered to be the origin of the strong magneto-crystalline anisotropy along the [001] direction. The spin-dependent band dispersions of $RuO_2$, especially along the Γ-M line around the Fermi level, which is a distinctive feature of altermagnetism and is consistent with the previous report.[6] Non-zero MAE contributions of the spin-conserving and spin-flip terms are also found along the Γ-M line as shown in Fig. 4f, which indicates that the MAE contribution will be enhanced in the Brillouin zone with the spin-polarized band dispersion. Therefore, the DFT calculations of single-variant RuO$_2$ indicate that the large MAE with antiferromagnetic order originates from the stability by forming spin quadrupole states along the Néel vector direction.

### D. Spin-splitting magnetoresistance in RuO$_2$(101)/CoFeB heterostructures

We further investigated the effect of this single variant of RuO$_2$(101) on the spin transport in the RuO$_2$(101)/CoFeB heterostructure. It is known that spin accumulation at the interface between a spin current source layer and a ferromagnetic layer alters the chemical potential of the ferromagnetic layer, which leads to a change in the resistance of the heterostructure when the spin-polarization (**s**) direction of the spin current is parallel (P) and antiparallel (AP) to the magnetization (**M**) direction of the ferromagnetic layer, the so-called unidirectional spin Hall magnetoresistance (USMR).[37] In the RuO$_2$(101) thin film, when the charge current (electric field, **E**) is applied in the [10$\bar{1}$] direction ($x$ direction), the generated spins are along the $y = z \times E$ direction, independent of the Néel vector.[13] The P and AP states for **M** and **s** can be achieved when the **M** of the CoFeB layer is aligned in the film plane at 0 and 180º to the **s**, as illustrated in Fig. 5a and 5b. We performed the second harmonic method[37] to measure the longitudinal resistance of a Hall bar device of the RuO$_2$(101)/CoFeB sample. We rotated the **M** of CoFeB in the (10$\bar{1}$) plane of RuO$_2$ under a magnetic field of 10 kOe, see the geometry in the top panel of Fig. 5c. The longitudinal resistance ($R_{2\omega}$) of RuO$_2$(10 nm)/CoFeB (2 nm) as a function of the rotation angle $\theta$ is shown in Fig. 5c. By fitting the curve, we find that it exhibits a sinusoidal dependence, i.e., $R_{2\omega}$ ~ $\sin\theta$, which is consistent with the USMR curves in Pt and Ta.[37] The lowest and highest resistances appear at $\theta = 90º$ and 270º, corresponding to the AP and P states, respectively.

Afterwards, we changed the charge current to the [010] direction. In this case, the **E** applied along the [010] direction produces a tilted spin current due to the spin-splitting effect.[13,15] The tilted spin current can contribute to the spin accumulation at the interface with its spin direction angled to the film surface. Thus, when the **M** is rotated in the (010) plane, the rotation angles that realize the P and AP states for **M** and **s** are different from the former case. The P and AP states of **M** and **s** are realized when the **M** is rotated to a certain angle outside the film surface, as shown in Figs. 5d and 5e. From the resistance versus the rotation angle shown in Fig. 5f, we can clearly find that there is an angular shift with respect to the data curve in Fig. 5c, with the lowest and highest resistances at 70º and 250º, respectively. To further understand the angular dependence, we fit the curve with a composite formula, i.e., $R_{2\omega} \sim a\sin(\theta + 35°) + b\sin\theta$. Here the first term gives rise to the spin current generation associated with the antiferromagnetic order, i.e., the Néel vector, while the second term results from the spin current generation from the unpolarized bands of the oxygen sites, which is not associated with the antiferromagnetic order. $a$ and $b$ are the parameters representing the magnitude of the contribution of these two terms to the total spin current. The red solid line in Fig. 5f shows a well-fitted result, with a ratio of $a$ and $b$ of ~1.4. The result of the angle shift for realizing the P and AP states is the evidence for the tilt spin current from the spin splitting effect. We refer to this effect as spin-splitting magnetoresistance (SSMR), a new member of the family of spin-splitting effects, which is the contribution of the single variant of $RuO_2$(101) to the spin transport. We note that the first harmonic $R_\omega$, as a linear response to current, is typically attributed to the combined action of both the spin-splitting effect and the inverse spin-splitting effect in this system, which may also exibit an angle shift linked to the Néel vector of $RuO_2$ in its angular dependence.

In conclusion, we present compelling evidence for the synthesis of single-variant altermagnetic $RuO_2$(101) thin films, epitaxially grown on $Al_2O_3$ ($1\bar{1}02$) substrates. Using atomic resolution transmission electron microscopy, we find the alignment of oxygen atoms at the $RuO_2$(101) and $Al_2O_3$($1\bar{1}02$) interfaces plays a critical role in single-variant formation. The unique ability of XMLD was utilized to detect the square of the element-specific magnetization proves essential for observing altermagnetism, which is hidden from conventional techniques due to its compensated magnetic order. Our single-variant films exhibit SSMR, making them potential for spin current generation in spintronic devices. The above results indicate the single variant features, while a multi-domain formation cannot explain such anisotropic properties. The DFT calculations also suggest the MAE with finite Ru magnetic moments. This work provides an intriguing correlation between altermagnetic features with magnetic spectroscopies and spin transport and may inspire further exploration for potential spintronic applications of altermagnetic materials.

Note added: During the review process, we became aware of relevant work reporting XMLD signals in (110)- and (100)-oriented $RuO_2$ films,[38] as well as SSMR results corresponding to the first harmonic resistance in (101)-oriented $RuO_2$/Co bilayers[39].

**METHODS**

**Material deposition and characterizations**

All the thin films were prepared using an ultra-high vacuum magnetron co-sputtering system with a base pressure of $6\times10^{-7}$ Pa. Pure Ru was used as the sputtering target. The $Al_2O_3(1\bar{1}02)$ substrates were treated in a muffle furnace at 1000 °C for one hour in an atmospheric environment. $RuO_2$(30 nm) films were grown on $Al_2O_3(1\bar{1}02)$ substrates by rf magnetron sputtering in a reactive sputtering atmosphere of Ar (30sccm) + $O_2$ (2.5 sccm) at 300 °C. Post-annealing in a high vacuum was performed to improve the crystallinity and resistivity. A combination of techniques was applied to characterize the films: *in-situ* reflection high-energy electron diffraction (RHEED), *ex-situ* atomic force microscopy (AFM), and out-of-plane X-ray diffraction (XRD) using Cu $K_{\alpha1}$ radiation. Further detailed microstructural characterization was performed using high-resolution high-angle annular dark-field scanning transmission electron microscopy (HAADF-STEM) with nano beam electron diffraction (NBED), and energy dispersive X-ray spectroscopy (EDS) on an FEI Titan G2 80–200 ChemiSTEM system.

**Measurement of X-ray magnetic spectroscopies**

The XAS, XMCD, and XMLD measurements were performed at BL-7A and 16A in the Photon Factory at the High-Energy Accelerator Research Organization (KEK). The total electron yield mode was adopted and all measurements were performed from 80 K to 400 K. For the XMCD measurements, the photon helicities of the incident beams were switched and defined as $\sigma^+$ and $\sigma^-$ spectra. The XMCD measurement geometries were set to 45° tilted from the normal incidence. In XMLD measurements, the direction of electric field component of incident synchrotron beam **E** is tuned horizontally and vertically with respect to the antiferromagnetic Néel vector direction **M**. We define the sign of the XMLD by subtracting (**M**∥**E**)−(**M**⊥**E**) spectra. For the angle dependences, the sample was rotated with the rotation axis of in-plane and out-of-plane of the sample.

**Device fabrication and transport measurements**

In the microfabrication, the films were patterned into bar devices (width: 10 μm, length: 25 μm) for second harmonic measurements. Electrodes consisting of a Ta (5 nm)/Au (100 nm) layer were then applied to the structures using sputtering followed by a lift-off process. The second harmonic measurements were conducted at room temperature with a Keithley 6221 and a lock-in amplifier LI 5660. Additionally, an external magnetic field of 10 kOe was applied and the sample was rotated as $\theta = 0 - 360°$. Transport properties of $RuO_2$ thin films deposited at various conditions were measured (Supplemental Materials). Spin-torque ferromagnetic resonance experiments were also performed to examine the generation of titled spin currents (see details in Supplemental Materials).

**First-principles calculations**

We investigated the MCA of rutile-type $RuO_2$ by the density-functional theory including the spin-orbit interactions (SOI), which is implemented in the Vienna *ab initio* simulation program (VASP).[40] We adopted the spin-polarized generalized gradient approximation (GGA)[41] for the exchange-correlation energy and used the projector augmented wave (PAW) potential[42,43] to treat the effect of core electrons properly. We considered the on-site Coulomb interaction[44] $U = 3$ eV for Ru atom, which is a necessary condition to obtain the altermagnetic state for $RuO_2$[21]. The lattice constants of rutile-type $RuO_2$ were used, $a = b = 0.4523$ nm and $c = 0.3115$ nm, which are obtained by a structure optimization with the VASP-PAW calculation. Furthermore, we performed second-order perturbation calculations of SOI for a more detailed understanding of the MCA mechanism. The theoretical details of the perturbation calculations can be found in previous works.[45,46] The convergence of these calculations was confirmed by 30×30×42 *k*-points in the Brillouin zone, which are sufficient to accurately estimate the MCA energy. For the calculation of the stability of the $RuO_2(101)$ and $RuO_2(\bar{1}01)$ variants by stacking $RuO_2(101)$ and $RuO_2(\bar{1}01)$ cells on the oxygen-terminated $Al_2O_3(1\bar{1}02)$ surface, a supercell containing 162 atoms (Al, O, and Ru) was used, with 4 monolayers of $RuO_2$ modeled as either $RuO_2(101)$ or $RuO_2(\bar{1}01)$.

## DATA AVAILABILITY

All data that support the findings of this study are included in the manuscript and supplementary materials. Source data (source data.zip) for figures are provided with the paper.

22. Han, L. *et al.* Electrical 180° switching of Néel vector in spin-splitting antiferromagnet. *Sci. Adv.* **10**, eadn0479 (2024).

23. Mazin, I. I., Koepernik, K., Johannes, M. D., González-Hernández, R. & Šmejkal, L. Prediction of unconventional magnetism in doped FeSb2. *Proc. Natl. Acad. Sci.* **118**, e2108924118 (2021).

24. Berlijn, T. *et al.* Itinerant Antiferromagnetism in RuO2. *Phys. Rev. Lett.* **118**, 077201 (2017).

25. Zhu, Z. H. *et al.* Anomalous Antiferromagnetism in Metallic RuO2 Determined by Resonant X-ray Scattering. *Phys. Rev. Lett.* **122**, 017202 (2019).

26. Feng, Z. *et al.* An anomalous Hall effect in altermagnetic ruthenium dioxide. *Nat. Electron.* **5**, 735–743 (2022).

27. Šmejkal, L., MacDonald, A. H., Sinova, J., Nakatsuji, S. & Jungwirth, T. Anomalous Hall antiferromagnets. *Nat. Rev. Mater.* **7**, 482–496 (2022).

28. Chi, B. *et al.* Crystal-facet-oriented altermagnets for detecting ferromagnetic and antiferromagnetic states by giant tunneling magnetoresistance. *Phys. Rev. Appl.* **21**, 034038 (2024).

29. Lin, T., Tomaz, M. A., Schwickert, M. M. & Harp, G. R. Structure and magnetic properties of Ru/Fe(001) multilayers. *Phys. Rev. B* **58**, 862–868 (1998).

30. Kim, D. H. *et al.* Correlation between Mn and Ru valence states and magnetic phases in SrMn1-xRuxO3. *Phys. Rev. B* **91**, 075113 (2015).

31. Wakabayashi, Y. K. *et al.* Isotropic orbital magnetic moments in magnetically anisotropic SrRuO3 films. *Phys. Rev. Mater.* **6**, 094402 (2022).
19

**ACKNOWLEDGMENTS**

This work was partially supported by the JSPS KAKENHI (Grant Nos. 20K04569, 20H00299, 21H01750, 22H04966, and 24H00408), MEXT Initiative to Establish Next-generation Novel Integrated Circuits Centers (X-NICS) Grant Number JPJ011438, Yazaki memorial foundation for science and technology, and the GIMRT Program of the Institute for Materials Research and the Cooperative Research Project Program of the Research Institute of Electrical Communication, Tohoku University. Dr. J. Uzuhashi and Dr. T. Furubayashi are acknowledged for their helps on TEM sample preparation and XRD measurements, respectively. C. He thanks Prof. K. Hono (NIMS) and Prof. S. Xu (SKL-ADMV, Hunan Univ.) for valuble suggestions and discussions.

**AUTHOR CONTRIBUTIONS**

Z.W., J.O., and C.H. conceived and designed the research. S.M. supervised the study. C.H. deposited the thin films and carried out RHEED, AFM, and STEM observations. Z.W. performed XRD measurements, device fabrication, and second-harmonic measurements. J.O. performed XMLD measurements. Y.M. carried out the first principles calculations. C.H., Z.W. and J.O performed data analysis with contributions from S.M., H.S., T.M., T.O., T.S., and Y.M.. C.H., Z.W., J.O., and Y.M. wrote and revised the manuscript with input and comments from all authors.

**COMPETING INTERESTS**

The authors declare no competing interests.

**Figures and Captions**

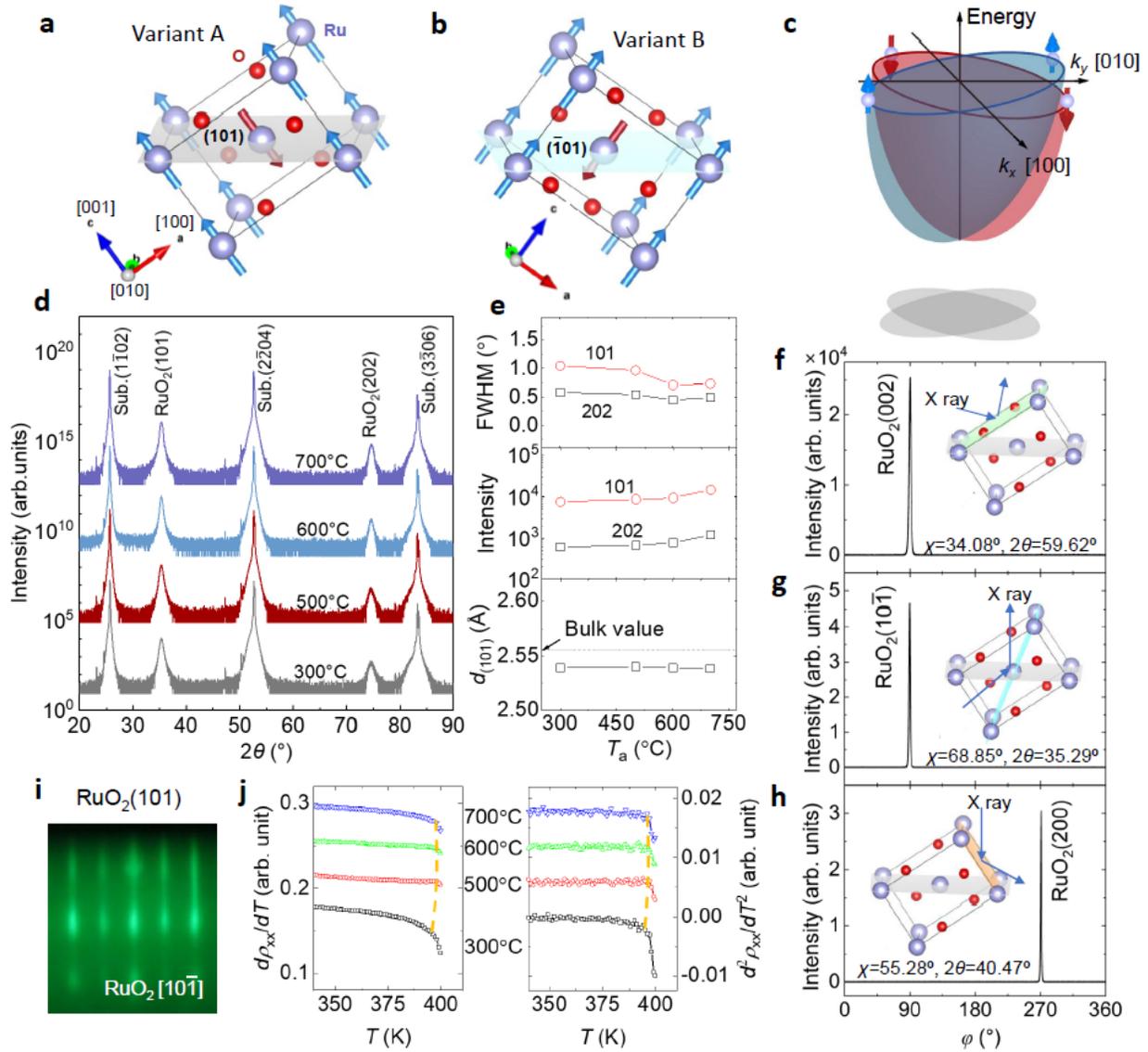

**Figure 1. Illustration of altermagnetic RuO₂ and structural and transport properties. a, b**, Illustrations of RuO₂ crystal structure labeled with spin directions and (**a**) Variant A with (101)

lattice plane and (**b**) Variant B with ($\bar{1}$01) lattice plane. **c**, Spin-splitting Fermi surface of the altermagnetic $RuO_2$. **d**, **e**, Out-of-plane XRD analysis of the $RuO_2$ thin films grown on the *r*-plane sapphire substrate at 300 °C and then annealed at $T_a$ = 300, 500, 600 or 700 °C for 15 min. **f-h**, In-plane XRD (*φ*-scan) patterns of the (002), (10$\bar{1}$), and (200) planes of the $RuO_2$(101) film. **i**, RHEED pattern of the film after the annealing at 600 °C. **j**, First and second derivative of resistivity as a function of measuring temperature for the $RuO_2$ thin films with different $T_a$. The dashed lines indicate the continuity changes corresponding the Néel temperature of the antiferromagnetic-paramagnetic transition.

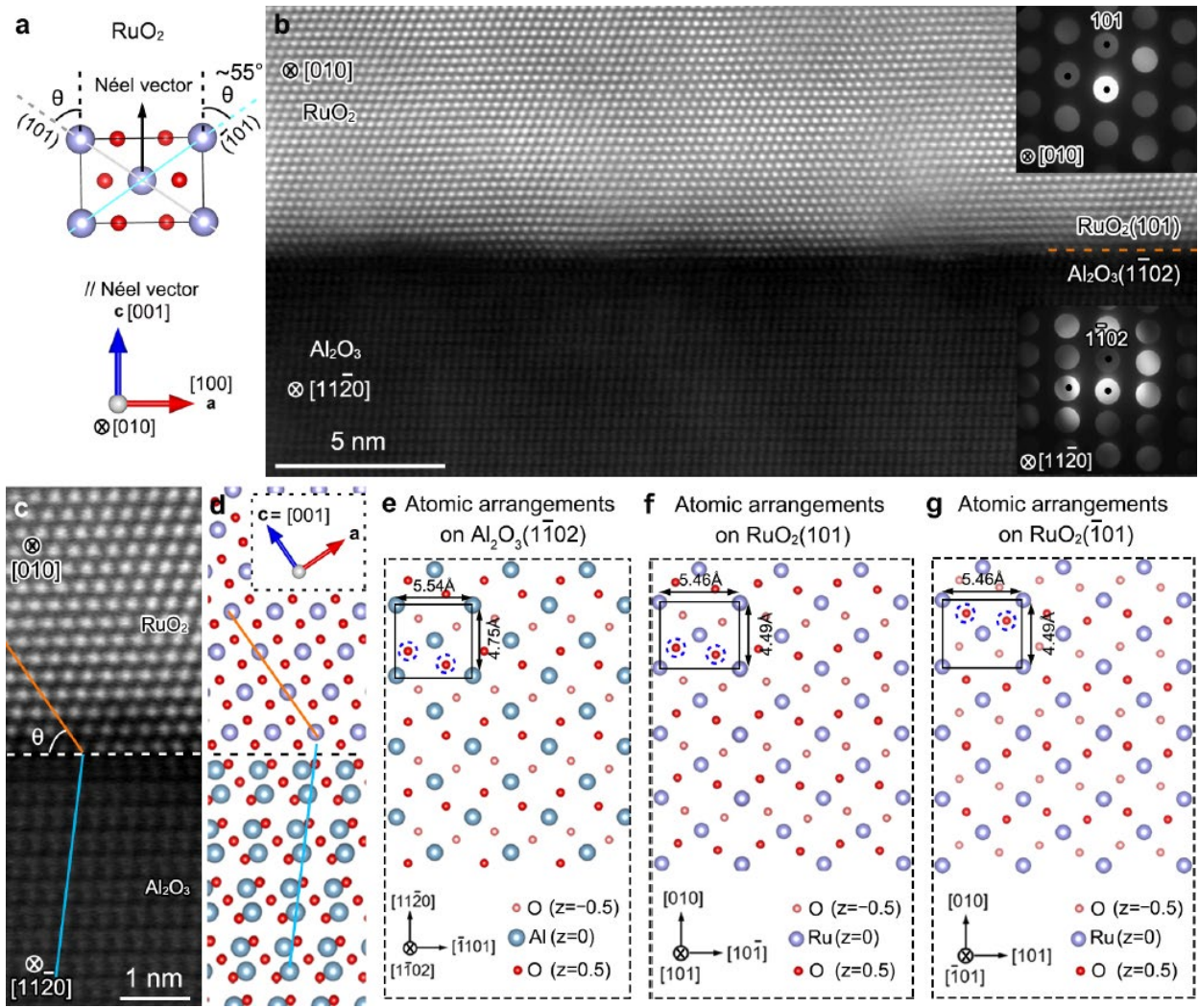

**Figure 2. Atomic-resolution STEM observations. a**, Schematic diagram showing the geometric relationship among the Néel vector, $RuO_2$(101) and ($\bar{1}$01) planes. **b**, HAADF-STEM image

showing the cross-sectional microstructure of RuO$_2$ film grown on *r*-plane Al$_2$O$_3$ single crystalline substrate. The upper-right and lower-right insets show the NBED patterns of RuO$_2$ and Al$_2$O$_3$, collected from RuO$_2$[010] and Al$_2$O$_3$[11$\bar{2}$0], respectively. **c**, Enlarged image showing the atomic arrangements at the interface between RuO$_2$(101) and Al$_2$O$_3$(1$\bar{1}$02). **d**, Schematic diagram of the lattice match between RuO$_2$ and Al$_2$O$_3$, corresponding to the HAADF image in (**c**). **e-g**, Schematic diagrams of the in-plane atomic arrangements in Al$_2$O$_3$(1$\bar{1}$02), RuO$_2$(101) and RuO$_2$($\bar{1}$01), respectively. The dashed blue circles in the black boxes highlight the positions of O atoms.

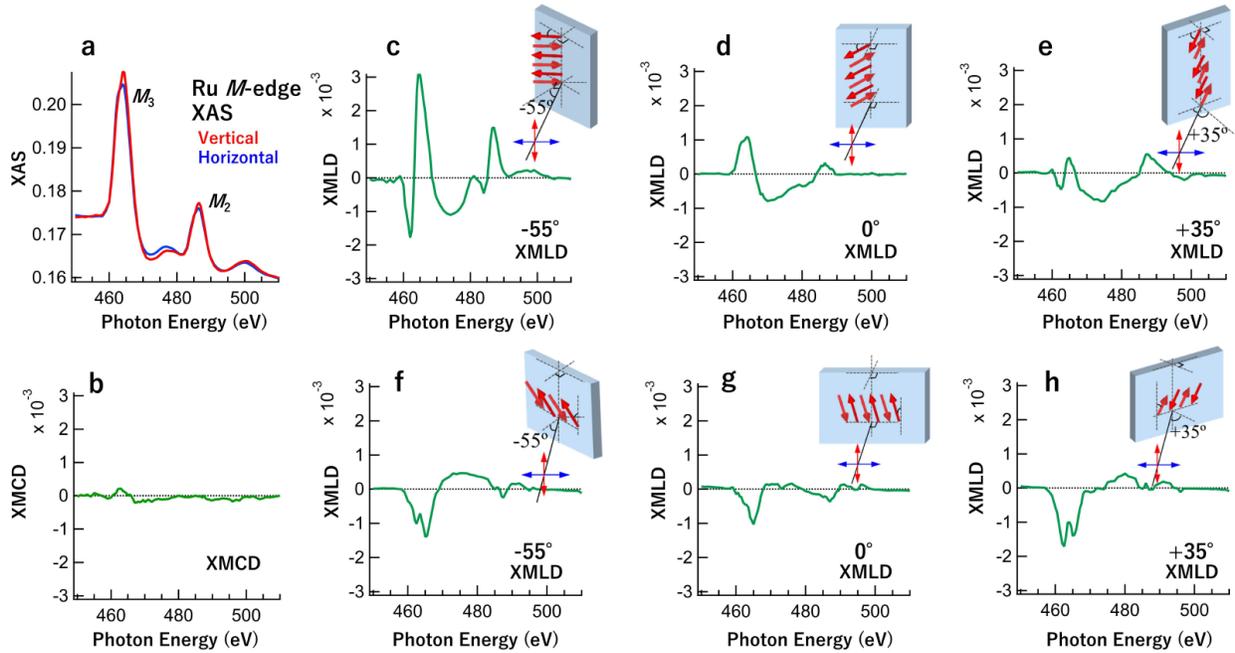

**Figure 3. X-ray magnetic spectroscopies with angular dependence in RuO$_2$(101) thin films.**
**a**, XAS taken by linearly polarized beams at $\theta = -55°$. **b**, XMCD by the difference of circularly polarized beams. **c-h**, XMLD with angular dependence. In the top three panels, the sample was rotated along the in-plane axis of $\theta = -55°$, 0°, and +35° as displayed in the illustrations. The directions of the Néel vectors in RuO$_2$ are also illustrated (red arrows). The bottom three panels display the cases of sample rotation $\varphi = 90°$ along the sample surface normal. Vertical axis scales are unified for all XMCD and XMLD panels. All measurements were performed at 80 K.

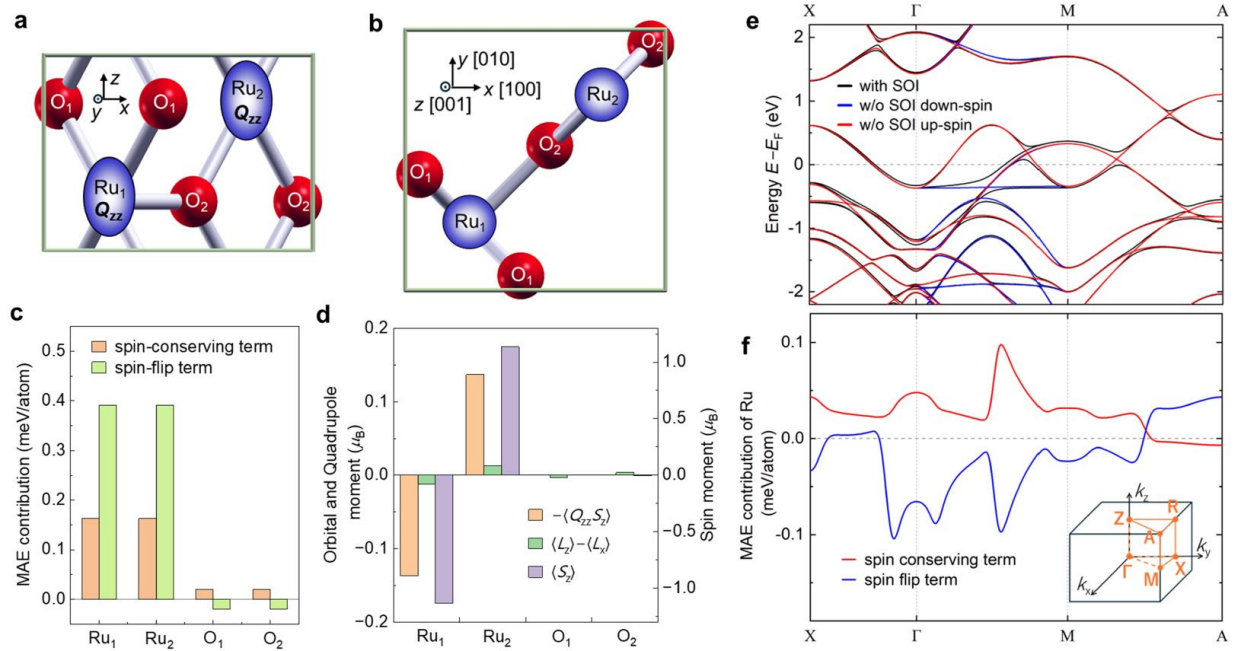

**Figure 4. Magnetic calculations for altermagnetic RuO₂. a,b**, Schematic illustration of the sites of Ru and O in $RuO_2$, as well as the quadrupole and spin moments for the perpendicular magnetization from (**a**) the y-direction and (**b**) the z-direction. x, y, and z indicate the [100], [010], and [001] directions, respectively. **c**, MAE contribution of each atom by the second order perturbation calculation in $RuO_2$. **d**, Spin quadrupole $-\langle Q_{zz}S_z \rangle$ and orbital moment anisotropy $\langle L_z \rangle - \langle L_x \rangle$ of Ru and O atoms in $RuO_2$. **e**, Band dispersion with and without spin orbit interaction (SOI) of $RuO_2$ along high symmetry lines. **f**, The MAE contribution of the Ru atoms by the second order perturbation calculation along the high symmetry line along $X(0, \frac{1}{2}, 0)$, $\Gamma(0, 0, 0)$, $M(\frac{1}{2}, \frac{1}{2}, 0)$, and $A(\frac{1}{2}, \frac{1}{2}, \frac{1}{2})$. Inset shows the first Brillouin zone with the *k* paths and the high symmetry lines.

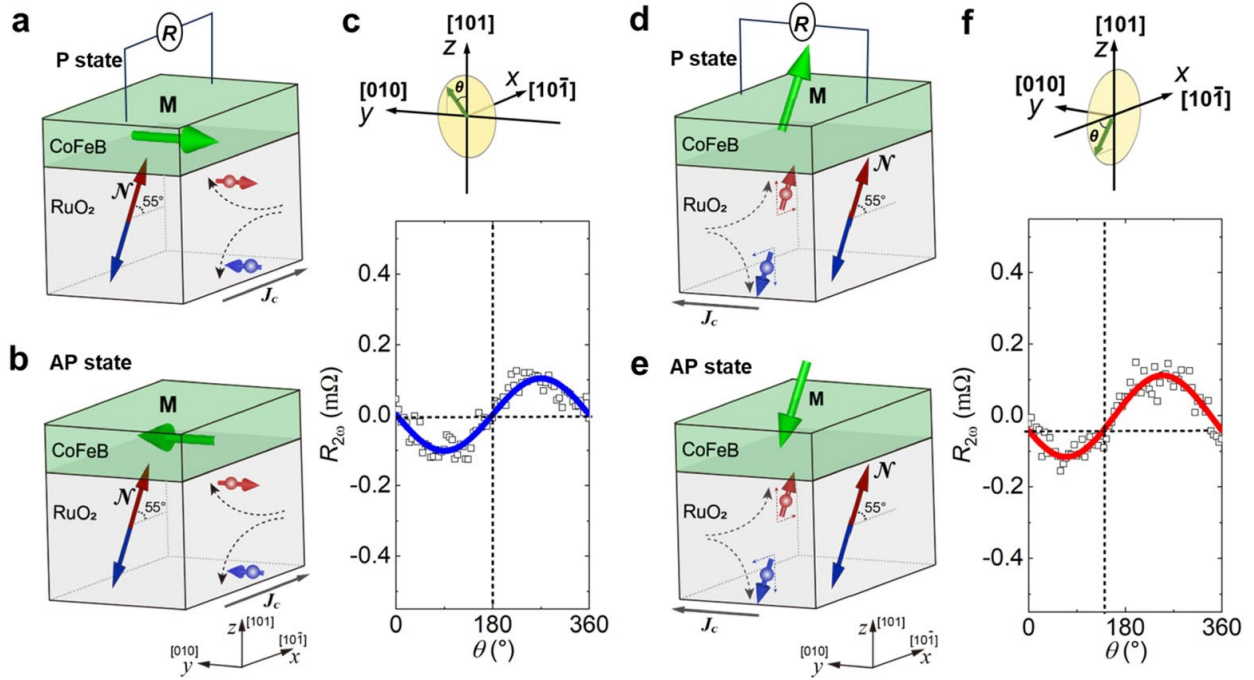

**Figure 5. SSMR in the RuO$_2$/CoFeB bilayers. a**, **b**, Illustration of spin current generation when the **E** is along the [10$\bar{1}$] direction, and **M** and **s** are in (**a**) P and (**b**) AP states. **c**, Angular dependence of $R_{2\omega}$ of RuO$_2$(101)/CoFeB bilayers measured with rotating the **M** in the (10$\bar{1}$) plane. **d**, **e**, Illustration of tilted spin current when the **E** is applied in the [010] direction. **M** and **s** have (**d**) P and (**e**) AP states. **f**, $R_{2\omega}$ as a function of the rotation angle of **M** in the (010) plane. The solid lines are the fitting results.

# Supplementary Materials

## Evidence for single variant in altermagnetic RuO$_2$(101) thin films

Cong He,[1,§,#] Zhenchao Wen,[1,§,*] Jun Okabayashi,[2,*] Yoshio Miura,[1,3,*] Tianyi Ma,[1] Tadakatsu Ohkubo,[1] Takeshi Seki,[4,5] Hiroaki Sukegawa,[1] and Seiji Mitani[1]

[1]National Institute for Materials Science (NIMS), Tsukuba 305-0047, Japan

[2]Research Center for Spectrochemistry, The University of Tokyo, Bunkyo, Tokyo 113-0033, Japan

[3]Faculty of Electrical Engineering and Electronics, Kyoto Institute of Technology, Kyoto 606-8585, Japan

[4]Institute for Materials Research, Tohoku University, Sendai 980-8577, Japan

[5]Center for Science and Innovation in Spintronics, Tohoku University, Sendai 980-8577, Japan

[#]Present address: Hunan University, Changsha 410082, China

[§]These authors contributed equally to this work.

[*] Wen.Zhenchao@nims.go.jp, jun@chem.s.u-tokyo.ac.jp, miura@kit.ac.jp

## 1. Annealing temperature dependence of surface morphology and crystal structure

Figures S1a-c show the surface morphology of the RuO$_2$ (30 nm) thin films annealed at $T_a$ = 300 ºC, 500 ºC and 700 ºC, as characterized by atomic force microscopy (AFM). The corresponding average roughness ($R_a$) and peak-to-valley ($P$-$V$) values are summarized in Figures S1d and 1e, respectively. The $R_a$ increases slightly to 0.30 nm as $T_a$ is raised to 600 ºC, and then increases dramatically to 0.54 nm with a $P$-$V$ value of 5.30 nm. Figures S1f-h display the reflection high-energy electron diffraction (RHEED) patterns for the samples annealed at the above temperatures. Clearly discernible RHEED streaks are observed for each sample. Look further

into the details, these streaks become progressively sharper with increasing $T_a$, indicating the improved crystallinity at higher annealing temperatures.

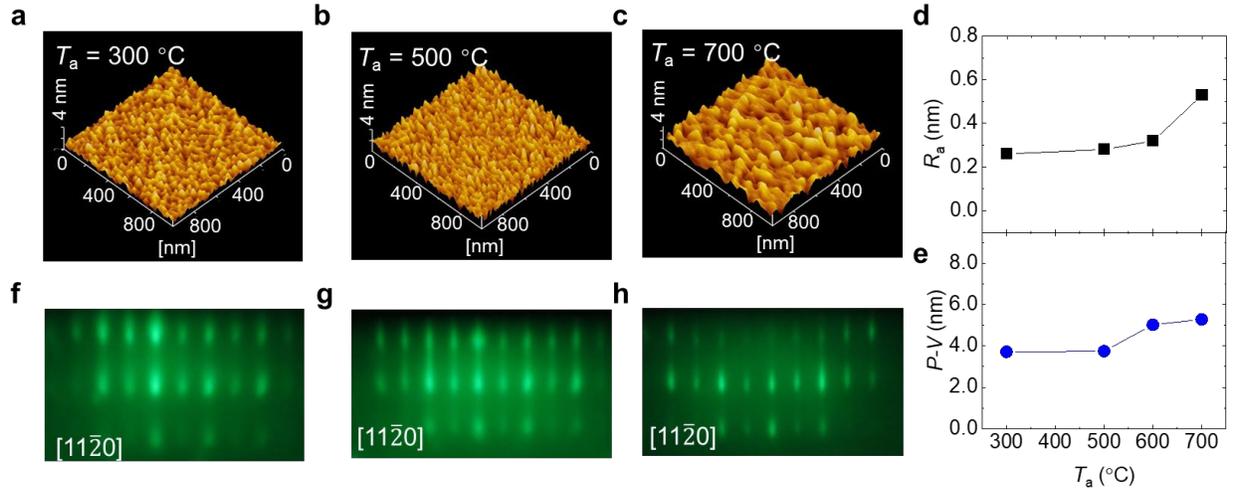

**Figure S1. Surface morphology and crystalline structures of RuO$_2$ thin films. a-c**, AFM images of a 30-nm-thick RuO$_2$ thin film annealed at $T_a$ = 300 ºC, 500 ºC and 700 ºC. **d, e**, $T_a$ dependence of $R_a$ and $P$-$V$ values for the 30-nm-thick RuO$_2$ sample. **f-h**, RHEED patterns of the RuO$_2$ thin film at $T_a$ = 300 ºC, 500 ºC and 700 ºC, respectively. The incident electron beam is along the azimuth of Al$_2$O$_3$[11$\bar{2}$0].

## 2. Transport properties of RuO$_2$ thin films deposited at various conditions

We prepared RuO$_2$ thin films on different substrates inlcuding $r$-plane Al$_2$O$_3$(1$\bar{1}$02), $c$-plane Al$_2$O$_3$(0001), and MgO(001) substrates with different deposition temperatures and oxygen flow rates. Fig. S2a shows the annealing temperature dependence of conductivity of the RuO$_2$ thin films for different growth conditions. Note that the unspecified growth conditions are a growth temperature of 300 ºC and an O$_2$ ratio of 7.7% in Ar and O$_2$ mixture. We found that the RuO$_2$ films grown on the Al$_2$O$_3$(1$\bar{1}$02) substrate had the largest conductivity. The conductivity increases as the annealing temperature increases and a maximum conductivity of $2.1 \times 10^6$ $\Omega^{-1}$m$^{-1}$ appears at an annealing temperature of $T_a$ = 600 ºC. Figure S2b shows the resistivity of the RuO$_2$ films grown on the Al$_2$O$_3$(1$\bar{1}$02) substrate as a function of measured temperature. It was found that the resistivity decreases with temperature decreases, showing a clear metallic behavior. Figure S2c summarizes the annealing temperature dependence of the residual resistivity as well as the residual-resistance ratio (RRR). The RuO$_2$ films with high annealing temperatures have small residual resistivities and high RRR, which could be due to the formation of better crystallinity at higher annealing temperatures, in agreement with the results of xrd and RHEED observations. In addition, we also performed the first-order and second-order differentiation of the curves of resistivity versus measurement temperature to confirm the Néel temperature of the RuO$_2$ films, as shown in Fig. 1j in main text. the continuity of the differential curves change was clearly observed, which could corresponds to the transition between antiferromagnetic order and paramagnetic order, i.e., the Néel temperature. We found that for the samples annealed at $T_a$ = 600 ºC and 700 ºC, the Néel temperature is around 390 K, comparable to the previous report.[1]

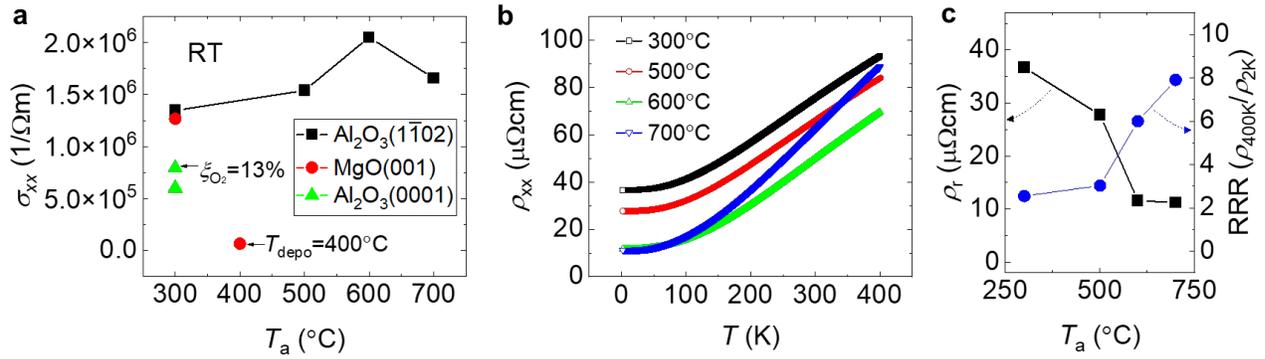

**Figure S2. Annealing and measuring temperature dependences of electric transport properties of the RuO$_2$ thin films deposited at various conditions. a**, Conductivity as a function of annealing temperature $T_a$. for the RuO$_2$ thin films deposited on different substrates. **b**, Resistivity of RuO$_2$ thin films deposited on the *r*-plane Al$_2$O$_3$(1$\bar{1}$02) substrates with different $T_a$. **c**, $T_a$ dependence of residual resistivity and RRR for the Al$_2$O$_3$(1$\bar{1}$02) substrate//RuO$_2$ (30 nm) thin films.

### 3. Crystallographic analysis of crystal domains within RuO$_2$ film

Figure S3 compares the theoretical and experimental scenarios when RuO$_2$ film is deposited on *r*-plane sapphire Al$_2$O$_3$(1$\bar{1}$02) substrate. Figures S3a and 3c illustrate the lattice matching of RuO$_2$(101) and RuO$_2$($\bar{1}$01) grown on the Al$_2$O$_3$(1$\bar{1}$02) substrates, respectively. Although these two variants share a mirror relationship, as indicated by the direction of their Néel vectors, no grain boundaries or phase boundaries are detected in our RuO$_2$ film, as shown in Fig. S3b and Fig. 2b. The experimental HAADF-STEM result is completely consistent with the case of RuO$_2$(101) grown on Al$_2$O$_3$(1$\bar{1}$02), which verifies the sing-variant growth of RuO$_2$(101) rather than RuO$_2$($\bar{1}$01).

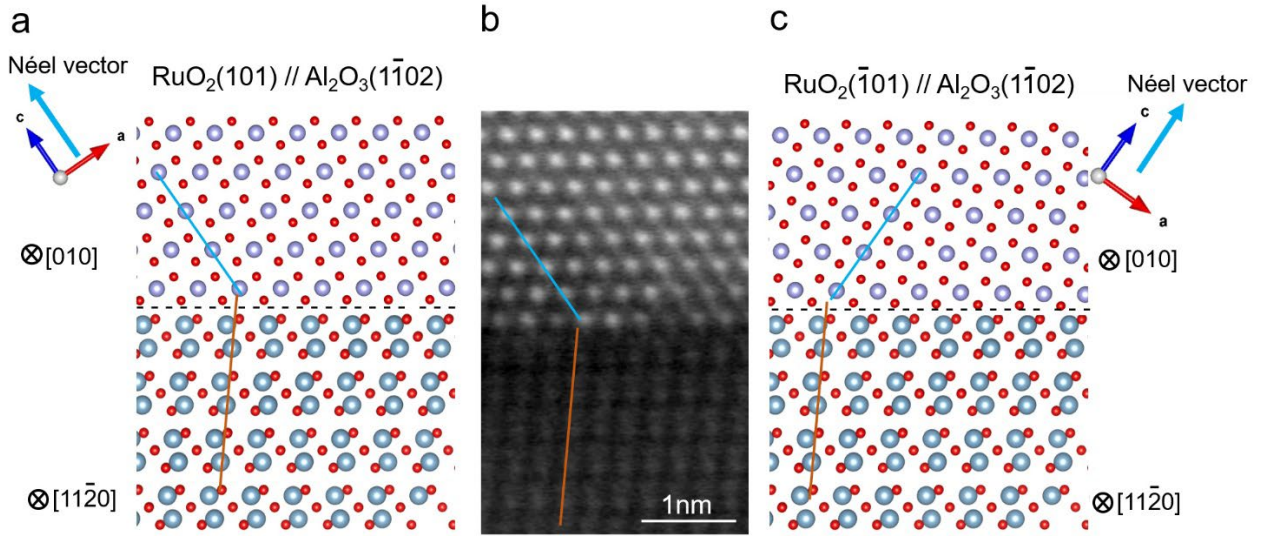

**Fig. S3. Crystallographic analysis of crystal domains within RuO$_2$ film deposited on the *r*-plane Al$_2$O$_3$(1$\bar{1}$02) substrate**. **a**, Schematic diagram showing the lattice matching for RuO$_2$(101) grown on the Al$_2$O$_3$(1$\bar{1}$02) substrate. **b**, HAADF-STEM image showing the direct experimental evidence of RuO$_2$ film grown on the Al$_2$O$_3$(1$\bar{1}$02) substrate. The light blue arrow represents the direction of Néel vector in RuO$_2$. **c**, Schematic diagram of the lattice matching between RuO$_2$($\bar{1}$01) and Al$_2$O$_3$(1$\bar{1}$02). The light blue and orange lines in (**a-c**) illustrate the atomic planes in the substrate and the film, respectively.

The low-magnification HAADF-STEM image in Fig. S4a further confirms that no multiple variants formed in the RuO$_2$ film, i.e. the successful single-variant growth of the film. EDS maps of different elements are shown in Figs. S4b-e. The quantitative line scan profiles in Fig. 4f show that the atomic ratios of O/Al and Ru/O in the substrate and the film are approximately 1.5 and 1.7, respectively. The former value is completely consistent with the stoichiometric ratio of $Al_2O_3$. The latter, however, is smaller than the theoretical stoichiometric ratio of $RuO_2$. This may suggest that there are a few O vacancies in the magnetron-sputtered $RuO_2$ film.

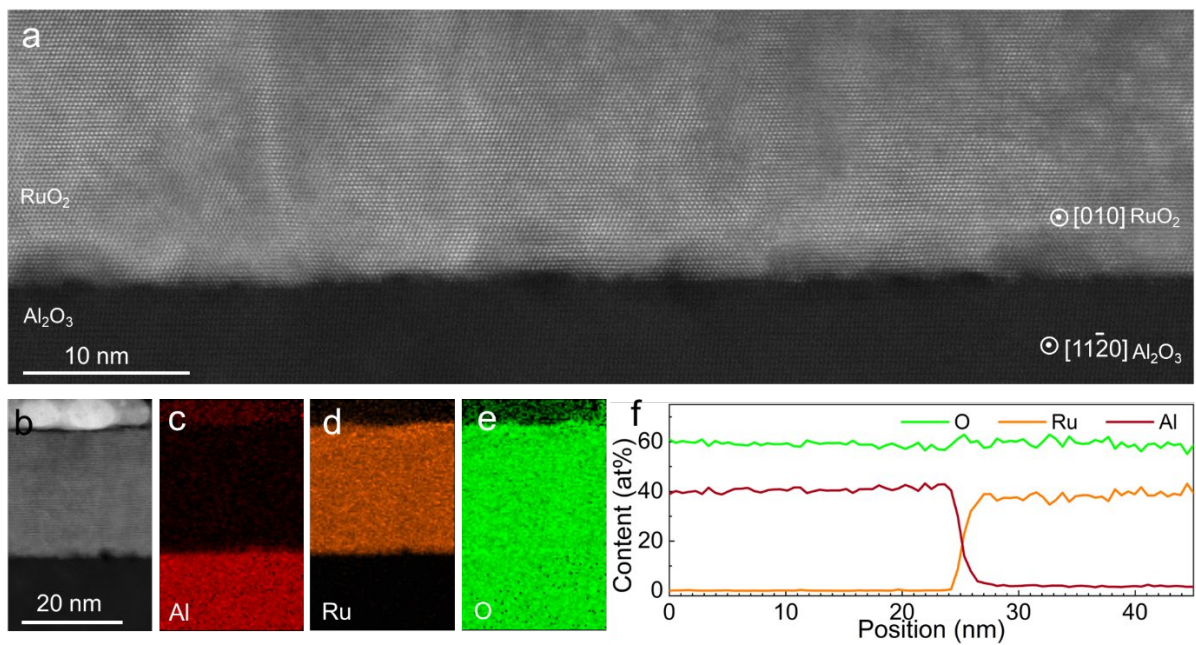

**Fig. S4. Chemical composition and O vacancy analyses in the $RuO_2$ film deposited on the r-plane of substrate, $Al_2O_3(1\bar{1}02)$. a**, Low-magnification HAADF-STEM image showing the single-crystalline growth on the *r*-plane $Al_2O_3(1\bar{1}02)$ substrate, with the electron beam direction of $Al_2O_3(11\bar{2}0)$. **b-e**, EDS mapping analysis of the film. **f**, The quantitative EDS line profiles of elements within the film.

## 4. The first-principles calculations for surface and variant stability

To clarify the stability of the $Al_2O_3(1\bar{1}02)$ surface, we conducted first-principles density functional theory (DFT) calculations using the VASP-PAW method. We optimized the surface structures of the $Al_2O_3(1\bar{1}02)$ for four different terminations, i.e., Type 1 and Type 2 with O and Al terminations, as shown in Fig. S5a-d. The Type 1 with O termination is corresponding to the Al$_2$O$_3$(1$\bar{1}$02) surface terminated with Al and O ($z = +0.5$) atoms while the Type 2 with O termination is for the Al$_2$O$_3$(1$\bar{1}$02) surface terminated with Al and O ($z = -0.5$) atoms. The calculations reveal that the oxygen-terminated Type 1 surface has the lowest surface energy density shown in Fig. S5 and is the most stable termination. In the experiments, we treated the Al$_2$O$_3$(1$\bar{1}$02) substrates in a muffle furnace at 1000 °C for one hour in an atmospheric environment. This preparation step also supports the formation of a stable oxygen-terminated surface.

We further calculated the stability of the RuO$_2$(101) and RuO$_2$($\bar{1}$01) variants by stacking RuO$_2$(101) and RuO$_2$($\bar{1}$01) cells on the oxygen-terminated Type-1 Al$_2$O$_3$(1$\bar{1}$02) surface, corresponding to Fig. 2f [Al$_2$O$_3$(1$\bar{1}$02)/RuO$_2$(101)] and Fig. 2g [Al$_2$O$_3$(1$\bar{1}$02)/RuO$_2$($\bar{1}$01)] in the manuscript. In these calculations, a supercell containing 162 atoms (Al, O, and Ru) was used, with 4 ML of RuO$_2$ modeled as either RuO$_2$(101) or RuO$_2$($\bar{1}$01). After structural relaxation, the cross-sectional atomic distributions in the two cases are shown in Fig. S6. We found that Al$_2$O$_3$(1$\bar{1}$02)/RuO$_2$(101) has a lower total formation energy density by 1.2 J/m$^2$ compared to Al$_2$O$_3$(1$\bar{1}$02)/RuO$_2$($\bar{1}$01). Here, the area density of formation energy is defined as $\gamma_{\text{formation}} = [E_{\text{total}} - (N_{\text{Ru}}\mu_{\text{Ru}} + N_{\text{Al}}\mu_{\text{Al}} + N_{\text{O}}\mu_{\text{O}})]/A$, where $N_{\text{Ru}}$, $N_{\text{Al}}$, and $N_{\text{O}}$ are the numbers of Ru, Al and O atoms, $\mu_{\text{Ru}}$, $\mu_{\text{Al}}$, and $\mu_{\text{O}}$ are the total energies per atom for bulk hcp-Ru, bulk fcc-Al, and O$_2$ molecule, and $A$ is the unit cell area of Al$_2$O$_3$(1$\bar{1}$02)/RuO$_2$(101). This result confirms that the Al$_2$O$_3$(1$\bar{1}$02)/RuO$_2$(101) is the more favorable configuration, aligning with experimental observations. Although the interface in the sample's TEM image is not perfectly sharp as a mono-atomic layer, we can find that, across the interface, there are two layers of Al atoms on the atomic terraces of the Al$_2$O$_3$(1$\bar{1}$02) substrate, which is consistent with the Type 1 oxygen-terminated case (Fig. S5a).

Overall, based on the DFT calculations, experimental TEM observations, and substrate preparation process, we conclude that the $Al_2O_3(1\bar{1}02)$ surface is Type 1 with oxygen-termination, allowing the growth of a single-variant $RuO_2(101)$ thin film. The interface structure and energy calculations further support that the $RuO_2(101)$ is a more stable and preferred orientation than $RuO_2(\bar{1}01)$.

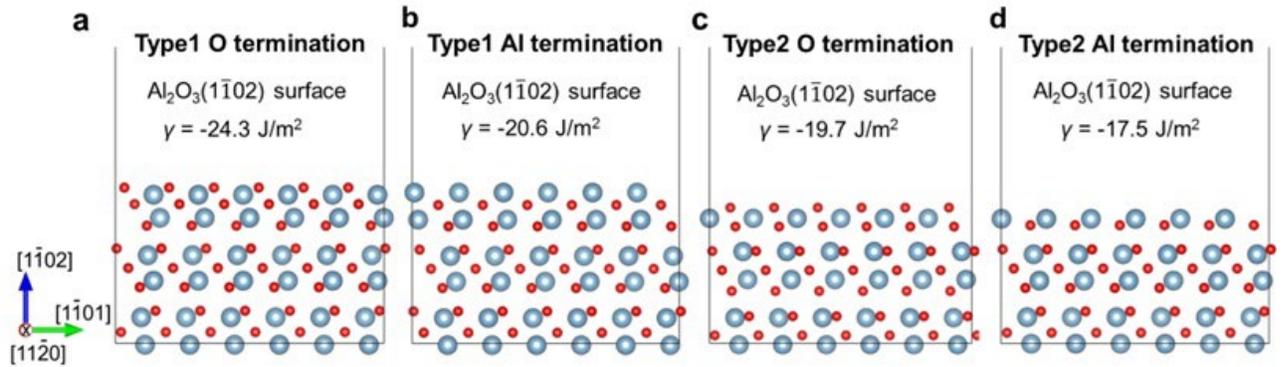

**Fig. S5**. Illustrations of structures of $Al_2O_3(1\bar{1}02)$ surface. **a**, Type 1 with O termination, **b**, Type 1 with Al termination, **c**, Type 2 with O termination, **d**, Type 2 with Al termination. Red and light blue dots represent oxygen and aluminum atoms, respectively. The formation energy density for each case is also shown.

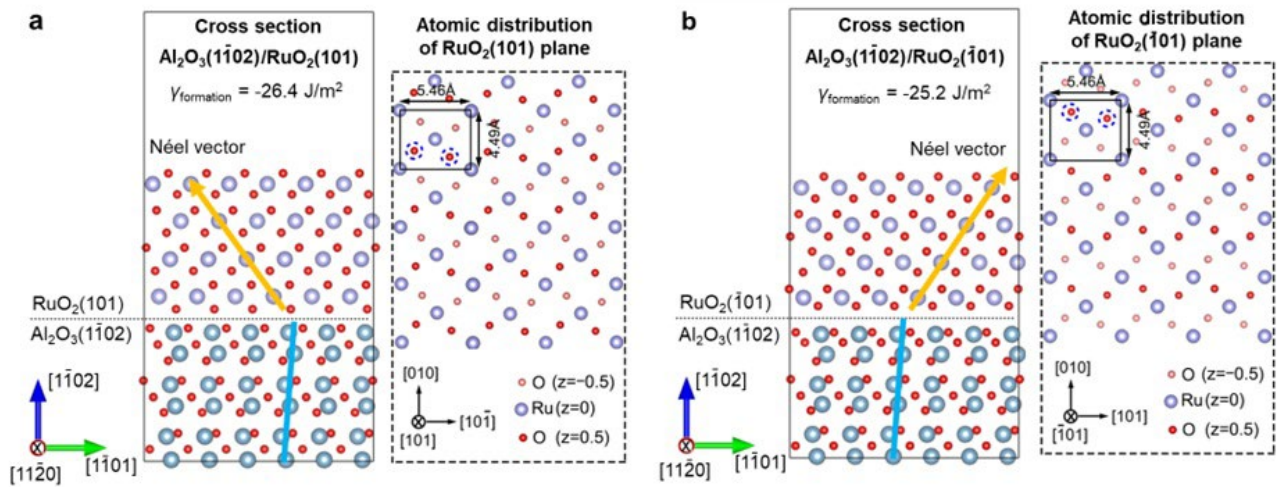

**Fig. S6**. Illustrations of cross-sectional and in-plane crystal structures and atomic distributions of **a**, Al$_2$O$_3$(1̄102)/RuO$_2$(101) and **b**, Al$_2$O$_3$(1̄102)/RuO$_2$(1̄01). Gray dots represent Ru atoms. Red and light blue dots are for oxygen and aluminum atoms, respectively. The formation energy density for each case is also shown.

## 5. Temperature dependence of XMLD, and integrals of XMLD intensities

Figure S5a displays the temperature-dependent XMLD spectra at 80, 200, 300, and 400 K. The spectra are normalized by XAS intensities. As the temperature increases, the intensity of the XMLD spectra diminishes, indicating a suppression of magnetic ordering. Given that the Neel temperature is approximately 400 K in RuO$_2$, the spectra at 400 K arise from the structural contribution, akin to X-ray linear dichroism (XLD). Consequently, we can distinguish between magnetic and structural contributions based on the temperature-dependent measurements. The contribution from the crystalline structure can be roughly estimated at 20% from Fig. S7a.

The integrals of XMLD intensities for $M_3$ and $M_2$ edges except for the 3$p$ to 4$s$ absorption regions are shown Fig. S7b. The XMLD sum rule can be derived as,

$$\langle Q_{zz} \rangle = \frac{l(2l-1)n_h}{2} \frac{\Delta I_{L3} + \Delta I_{L2}}{I_{L3} + I_{L2}}, \qquad (S1)$$

using the XAS and XMLD intensities of $I$ and $\Delta I$, respectively, for $L_2$ and $L_3$ edges.[2,3] Here, in the case of RuO$_2$, angular quantum number $l = 2$ for 4$d$ valence states and $n_h \sim 4$ for Ru$^{2+}$ (4d$^4$) can be adopted. $\langle Q_{zz} \rangle$ is estimated to be the order of 0.01, which results in the charge distribution of 1% order at the Ru sites.

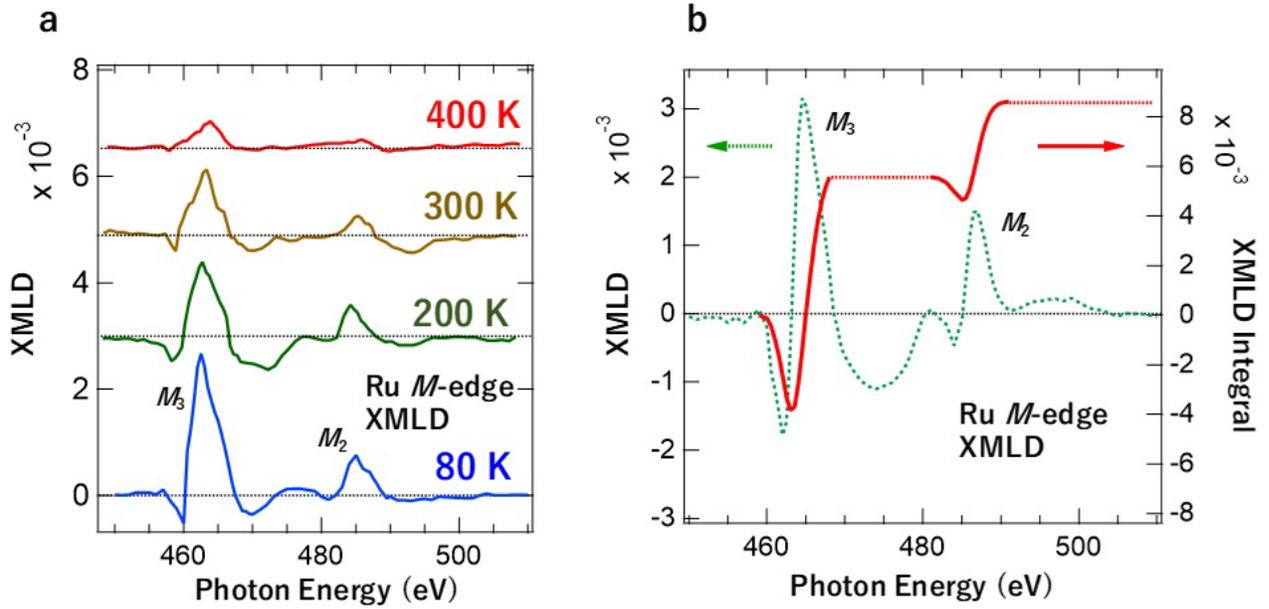

**Fig. S7. Analysis of XMLD spectral line shapes. a**, XMLD taken at 80, 200, 300, and 400 K. **b**, Integrals of XMLD intensities for $M_3$ and $M_2$ edges except for the $3p$ to $4s$ absorption regions.

## 6. Spin torque ferromagnetic resonance (ST-FMR) measurement in $RuO_2$(101)/CoFeB bilayers

We performed the ST-FMR measurement[4] to investigate the spin-orbit torques (SOTs) in the $RuO_2$(5 nm)/CoFeB (2 nm) bilayers deposited on the $Al_2O_3$($1\bar{1}02$) substrate. The film layers were patterned into devices with a rectangular shape separately along the $RuO_2$[$10\bar{1}$] and [010] directions with integrated Ta/Au electrodes as coplanar waveguides. Figures S8a and b show the illustrations of $RuO_2$/CoFeB bilayer structures with applying the charge currents along $RuO_2$[$10\bar{1}$] and [010] directions, as well as the generation of spin currents. The magnetic field was applied in plane of the film with the magnitude of 0 ~ 2500 Oe and the in-plane angle $\varphi$ of 0 ~ 360º. A radio frequency current was applied longitudinally to exert an oscillatory torque to induce precession in the CoFeB layer. This precession causes the device resistance to oscillate through the anisotropic magnetoresistance effect, producing a detectable mixed ST-FMR voltage $V$ that is measured by a lock-in amplifier. The ST-FMR voltage can be fitted by the following equation,[5]

$$V = V_S \frac{\Delta^2}{(H-H_{res})^2+\Delta^2} + V_A \frac{\Delta(H-H_{res})}{(H-H_{res})^2+\Delta^2} \qquad (S2)$$

where $\Delta$ is the resonance line width and $H_{res}$ is the resonance field. $V_S$ and $V_A$ correspond to the contributions of the damping-like (DL) and field-like (FL) SOTs. Typical FMR spectra for the RuO$_2$/CoFeB samples measured at $\varphi = 45°$ and frequency $f = 8$ GHz are shown in Figs. S6**c, d**. The fitting results using Eq. (S2) are also shown, and the symmetric $V_S$ and antisymmetric $V_A$ contributions of the Lorentzian line shapes of $V$ are separated. The in-plane angle $\varphi$ dependence of the ST-FMR signal was investigated, and the variations of the $V_S$ and $V_A$ components with respect to $\varphi$ are shown in Figs. S8 **e-h**. The DL and FL torques of $x$-, $y$-, and $z$-polarized spin currents can be separated from the angular dependence of $V_S$ and $V_A$ using the following equations,[6,7]

$$V_S(\varphi) \propto \sin2\varphi[\tau_x^{DL}\sin\varphi + \tau_y^{DL}\cos\varphi + \tau_z^{FL}], \qquad (S3)$$

$$V_A(\varphi) \propto \sin2\varphi[\tau_x^{FL}\sin\varphi + \tau_y^{FL}\cos\varphi + \tau_z^{DL}]. \qquad (S4)$$

Here $\tau_i^{DL}$ and $\tau_i^{FL}$ are the amplitudes of the DL and FL torques, respectively, where the $i$ indicates the $x$-, $y$-, and $z$-components of the spin currents. The fitting results of the angular dependence of the $V_A$ and $V_S$ of the ST-FMR signal are shown in Figs. S8**e-h**. It is found that for the device with the charge current flowing along the RuO$_2$[10$\bar{1}$] direction, $\sin2\varphi \times \cos\varphi$ dominates the fitting result, indicating that the torque originates mainly from spin current polarized in the $y$-direction, i.e., RuO$_2$[010] direction. However, for the device with the charge current flowing along the RuO$_2$[010] direction, the value of $\tau_z^{DL}$ is much larger than that of $\tau_y^{DL}$, indicating the generation of tilted spin currents.

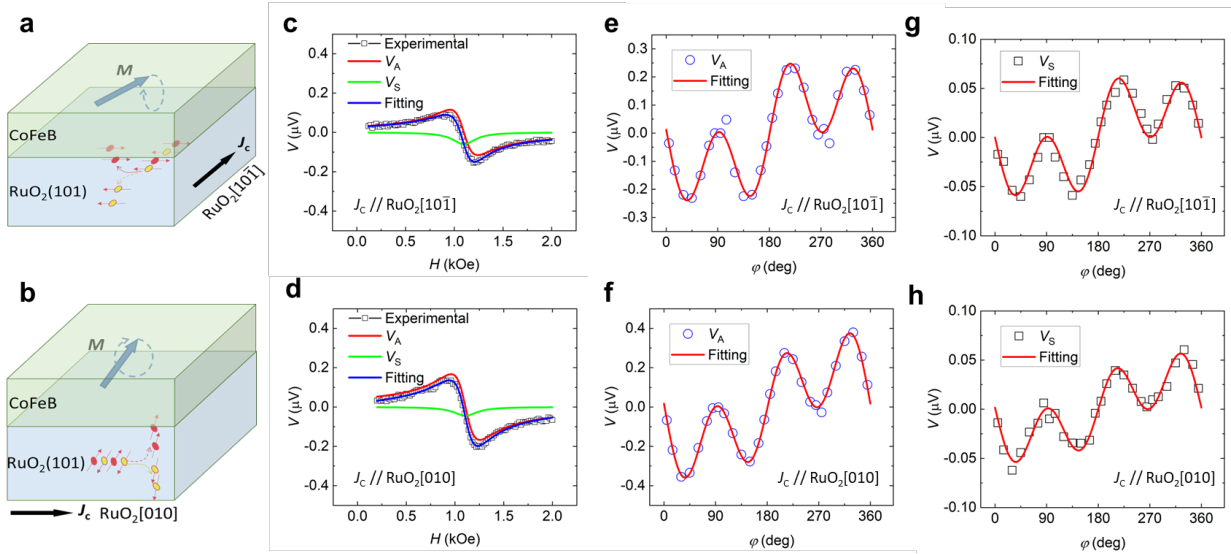

**Fig. S8. ST-FMR measurements with the charge current along the [10$\bar{1}$] and [010] directions. a, b,** Schematic illustrations of RuO$_2$(101)/CoFeB bilayer structures and spin current generation for the ST-FMR measurements with the charge current along the (**a**) [10$\bar{1}$] and (**b**) [010] directions, respectively. **c, d,** Typical ST-FMR spectra with fitting and the separation of antisymmetric ($V_A$) and symmetric ($V_S$) components. **e, f,** Antisymmetric voltage amplitudes depending on the angle of in-plane magnetic field. **g, h,** Symmetric component of voltage amplitudes as a function of in-plane magnetic field angle. Red solid lines are the fits using Eqs. (S3) and (S4). **c, e, g** are the results with the measurement configuration shown in (**a**), while **d, f, h** are with the configuration shown in (**b**).